\documentclass[useAMS,usenatbib]{mn2e}
\usepackage{graphicx}
\usepackage{amsmath}

\title[Ly$\alpha$ nebulae around high $z$ radio galaxies]{VIMOS-VLT spectroscopy of the giant Ly$\alpha$ 
nebulae associated with three $z\sim$2.5 radio galaxies.\thanks{Based on observations carried out at the
European Southern Observatory, Paranal (Chile). Programs 075.B-0212(A) and 073.B-0189(A).}}

\author[M. Villar-Mart\'\i n  et al.]
       {M. Villar-Mart\'\i n$^1$, S.F. S\'anchez$^2$, A. Humphrey$^3$, M. Dijkstra$^{4}$, S. di Serego Alighieri$^5$
\newauthor C. De Breuck$^6$, R. Gonz\'alez Delgado$^1$  \\
        $^{1}$Instituto de Astrof\'\i sica de Andaluc\'\i a (CSIC), Aptdo. 3004, 18080 Granada, Spain
(montse@iaa.es) \\
$^{2}$Centro Astron\'omico Hispano Aleman de Calar Alto (CSIC-MPIA), E4004 Almer\'\i a, Spain \\
$^{3}$Instituto de Astronom\'\i a, UNAM, Ap. 70-264, 04510 M\'exico, DF, M\'exico \\
$^{4}$School of Physics, University of Melbourne, Parkville, Victoria, 3010, Australia \\
$^{5}$INAF-Osservatorio Astrofisico di Arcetri, Largo Enrico Fermi 5, I-50125 Firenze, Italy\\
$^{6}$European Southern Observatory, Karl Schwarschild Str, 2, D-85748 Garching bei M\"unchen, Germany\\
}

\date{Accepted 2007 March 30.
      Received 2007 March 29; in original form 2006 November 17.}

\pagerange{\pageref{firstpage}--\pageref{lastpage}}
\pubyear{2007}

\begin{document}

\maketitle

\label{firstpage}

\begin{abstract}

The morphological and spectroscopic properties of the giant ($>$60 kpc) Ly$\alpha$ nebulae
 associated with three radio
galaxies at $z\sim$2.5 ( MRC 1558-003, MRC 2025-218 and MRC 0140-257)
have been investigated using  integral field spectroscopic data obtained with
VIMOS 
on VLT.

The morphologies are varied. The nebula of one source has a centrally
peaked, rounded appearance. In the other two objects, it consists of two spatial 
components. The three  nebulae are aligned with the radio axis
within $\la$30$^o$.
The total Ly$\alpha$ luminosities are in the range (0.3-3.4)$\times$10$^{44}$
erg s$^{-1}$.
The Ly$\alpha$ spectral profile shows strong variation through the nebulae, with FWHM
values in the range $\sim$400-1500 km s$^{-1}$ and velocity shifts $V_{offset}\sim$120-600 km s$^{-1}$.

We present an infall  model that can
explain successfully the 
morphology, size, surface brightness distribution and the velocity field of the Ly$\alpha$ nebula associated with MRC 1558-003. It can also explain why Ly$\alpha$ is redshifted
relative to other emission lines and the FWHM values of the non resonant HeII line. 
 This adds 
further
 support to our previous conclusion that the {\it quiescent} giant nebulae associated with
this and other high redshift powerful  radio galaxies are in  infall.
A problem for  this model  is the difficulty to reproduce the large Ly$\alpha$ FWHM
values, which might be consequence of a different mechanism. 

We have discovered a  giant ($\sim$85 kpc) 
Ly$\alpha$ nebula associated with the radio galaxy MRC 0140-257 at $z=$2.64.  It shows strikingly relaxed kinematics (FWHM$<$300 km s$^{-1}$ and $V_{offset}\la$120  km s$^{-1}$), 
unique among high $z$ ($\ga$2) radio galaxies. 

\end{abstract}


\section{Introduction}

Powerful high $z$ radio galaxies ($z\ga$2, HzRG) are often surrounded by giant Ly$\alpha$ nebulae which can extend for more than 100 kpc (e.g. McCarthy et al.  \citeyear{mac90a}; 
Reuland et al. \citeyear{reu03}, Villar-Mart\'\i n et al.  \citeyear{vm03}) and sometimes beyond the radio structures 
(e.g. Eales et al. \citeyear{eales93}; Kurk et al. \citeyear{kurk02}; Maxfield et al. \citeyear{max02}).
These are clumpy, irregular (with features such as filaments, plumes, ionization cones,
 e.g. Reuland et al. \citeyear{reu03}) and often aligned with the radio axis 
(McCarthy et al. \citeyear{mac95}). They are characterized by extreme kinematics, with measured FWHM  $\ga$1000 km s$^{-1}$ 
(e.g. McCarthy, Baum \& Spinrad \citeyear{mac96a}; Villar-Mart\'\i n et al.  \citeyear{vm03}), compared with values of a few hundred in low-redshift radio galaxies (e.g. Tadhunter, Fosbury \& Quinn \citeyear{tad89}; Baum, Heckman \& van Breugel\citeyear{baum90}). There is strong evidence  that interactions between the radio structures and the ambient
gas, produce an outflow  responsible for such extreme kinematics (e.g.  Humphrey et al. \citeyear{hum06}, van Ojik et al. 1997).

The nebulae have typical values of
ionized gas mass $\sim$10$^{9-10}$ M$_{\odot}$, Ly$\alpha$ luminosities
$\sim$several$\times$10$^{43-44}$ erg s$^{-1}$ and densities $n_e\sim$ few to several
hundred cm$^{-3}$ (e.g. McCarthy \citeyear{mac93}, Villar-Mart\'\i n et al. \citeyear{vm03}). 
They emit a rich  emission line spectrum  dominated in the optical
(UV rest frame) by Ly$\alpha$ followed by CIV$\lambda$1550, HeII$\lambda$1640 and
CIII]$\lambda$1909 (CIV, HeII and CIII] hereafter). Such a line spectrum reveals high levels
of metal enrichment  and excitation mechanisms 
mostly related to the nuclear activity, at least in the
direction along the
radio structures  (e.g. Vernet et al. \citeyear{ver01}).

In addition to the highly perturbed gas,
HzRG are often  embedded in giant (often $\ge$100
kpc), low surface brightness   nebulae  of metal rich, ionized gas
with {\it quiescent kinematics}  (Villar-Mart\'\i n et al. \citeyear{vm03}, Reuland et al. \citeyear{reu07}), i.e.,
not  perturbed by 
interactions with the radio structures. 
We have recently shown that these quiescent nebulae are
infalling towards the central region (Humphrey et al. \citeyear{hum07}).

There are only several narrow band Ly$\alpha$ images of HzRG 
and most spectroscopic studies have been performed with  the long-slit technique,
 with the slit aligned with the radio structures.
These studies have been  seriously limited by, respectively, the lack
of spectral information, or the lack of spatial information in directions other than the radio axis. For this
reason, we are carrying out an observational program
of 3D integral field spectroscopy of powerful radio galaxies at $z\sim$2-3
with VIMOS on VLT and PMAS/PPAK on the 3.5m telescope in Calar Alto Observatory. The
main goal is
 to characterize
the morphological, kinematic and ionization properties  of the extended 
 ionized gas in two spatial
dimensions.

In this paper, we present  results obtained for  MRC 1558-003 ($z$=2.53), MRC 2025-218 ($z$=2.63) and MRC 0140-257
($z$=2.64), based on VIMOS-VLT  data. Results on MRC 2104-242 ($z$=2.49) can be found in  \cite{vm06}.
A similar study of 4C40.36 ($z$=2.27) and 4C48.48 ($z$=2.34)
based on PMAS/PPAK data (3.5m telescope, Calar Alto Observatory) will be presented in S\'anchez et al. (2007, in prep.).

A  $\Omega_{\Lambda} =$ 0.73, $\Omega_{m}$ = 0.27 and $H_{0}$ = 62 km  $s^{-1}$  Mpc$^{-1}$ 
cosmology is adopted in this paper (Sandage et al. \citeyear{san06}).

\section{Observations and data reduction}

The observations (program 075.B-0212(A))  were made on UT 2005  July 28, 29 and 30 using the
 VIsible MultiObject Spectrograph (VIMOS, Le F\'evre et al. \citeyear{lefe03}),  on the Nasmyth focus
of the UT3 VLT. The instrument  is equipped with an integral field unit  with 6400  microlenses coupled to fibres. For the  configuration selected by us, the number of fibers
in use is 1600, covering 27'' x 27'' on the sky with a  0.67" sampling.  
The HR$_{\rm blue}$ grating was used, with  an effective wavelength range
 $\sim$4150-6200 \AA, an instrumental profile of FWHM 1.7$\pm$0.2 \AA\ and a pixel scale of
0.5 \AA\ pixel$^{-1}$.

The  exposure time on each target was 7.3 h (22$\times$1200 s) on MRC 1558-003, 10 h (30$\times$1200 s) on MRC 2025-218 and 8.3 h (25$\times$1200 s) on MRC 0140-257. 
In addition, we had 4.5 h (9$\times$1800 s) on MRC 1558-003 obtained in June 2004 (program 073.B-0189(A), see 
Villar-Mart\'\i n et al. \citeyear{vm06} for a description of these observations). The total integration time on this source  was therefore 11.8 h.

The seeing full width at half maximum during the observations was in the range $\sim$0.44-1.32'' (1st night),
0.4-1.2'' 2nd night) and 0.5''-3.0'' (3rd night) for the 2005 observations, and 1.0-1.4$\arcsec$
for the 2004 data on MRC 1558-003.

For each galaxy a dithering pattern was applied, with a maximum offset of $\sim$3'', and a range of dithering pointings between 3 and 7. 

The data were reduced using  R3D (S\'anchez 2006) and IRAF routines.
 The data were bias subtracted. 
The  locations of the spectra were traced on a continuum-lamp exposure
obtained before each target exposure. The corresponding spectrum was then extracted for each fiber  by
coadding the flux intensity within a 5 pixel aperture, along the spectral dispersion axis, centered on the
estimated location of the fiber centroid.

 The wavelength calibration was performed using  arc lamp spectra
 and
the telluric emission lines in the science data.   The wavelength solution
was stable within a fraction of 1 spectral pixel (0.5 \AA)  across the whole spectral range
and the whole field of view.

The wavelength calibration
in the blue (i.e. the Ly$\alpha$ spectral region) was problematic due to the lack of 
bright emission lines in the arc lamp and sky spectra. Residual, artificial  shifts in $\lambda$ 
 of up to 6 \AA\ were measured for Ly$\alpha$ relative to CIV and HeII in MRC 1558-003 and
MRC 2025-218. 
  We have been able to correct for this effect by comparing with available Keck long-slit spectroscopy
 (Villar-Mart\'\i n et al. \citeyear{vm03}, Humphrey \citeyear{hum04}),
to an accuracy of $\la$1 \AA. In any case, such uncertainty will not affect the results presented
here.

The fibre-to-fibre response  at each  wavelength
 was determined from a continuum-lamp exposure. 

After
these basic reduction steps, a data cube was created for each exposure. The cubes
for MRC 2025-218
 were then recentered spatially  at
each wavelength by determining the centroid of a nearby star in the
VIMOS field of view. This recentering corrects for
differential atmospheric refraction.

 For the other two objects, there are no stars     in the VIMOS field of view. We 
used the information on the spatial shifts contained in the image headers to 
apply the corresponding shifts. However, we found that this
technique left substantial spatial offsets. For this reason,
 we used the Ly$\alpha$ peak of emission to recenter all cubes. Although this recentering is valid in
the blue, it is not clear  that it is also valid
in the red. In particular, in the case of MRC 1558-003, 
we found a residual  spatial shift of $\sim$1 spaxel in the direction of the
radio structures between the blue (Ly$\alpha$) 
and red (CIV, HeII) parts of the spectrum, which is much larger than  found in previous works
(Villar-Mart\'\i n et al. \citeyear{vm99}; Humphrey et al. \citeyear{hum07}). 
For this reason, an additional correction was applied. Uncertainties
remain regarding the spatial centering in  the  direction perpendicular to
the radio axis, although this is likely to be $<$1 spaxel or 0.67$\arcsec$. This will not affect
seriously the results presented here.

For MRC 0140-257, no useful previous works or additional data were available. 
Although the accuracy  of the spatial centering between
the red and the blue ends of the spectrum is uncertain, this will not affect
the results presented here, due to the nature of our analysis and the clear separation between the 
spatial components of the Ly$\alpha$ nebula.

 The cubes were then combined using IRAF tasks, masking the broken and/or low sensitivity
fibres.  A 3$\sigma$ clipping algorithm removed cosmic
rays.
The sky background was estimated before subtraction by selecting the
spectra of object free areas and creating an interpolated datacube (using E3D,
S\'anchez \citeyear{san04}). A spatial median
smoothing using  a  4x4 spaxel box was applied.

The nights were non-photometric. Observations of
standard stars were used to perform a relative calibration from blue to red. 
The absolute flux calibration for MRC 1558-003 and MRC 2025-218
was done using   available Keck long-slit spectroscopy along the
radio axis. We extracted a pseudo-slit from the Vimos data as similar as possible
to the slit aperture of the Keck spectra. 
 The flux
in different spectral windows was then compared and scaled. The final flux calibration
has an accuracy of $\sim$20\%.
For MRC 0140-257, this could not be done, due to the unavailability of a high
quality long-slit spectrum,  but as we  explain in \S5.3, the agreement between the measured
Ly$\alpha$ flux and  published values is reasonably good ($\sim$20\%).

  Cross-talk effects are
estimated to be negligible.  The fibre to fibre contamination is expected to be  $<$5\% for adjacent spectra in the CCD, dropping to less than 1\% for the 3rd adjacent spectra. 

To  overlay the radio maps   (Carilli et al. \citeyear{car97}, Pentericci et al. \citeyear{pente00b}) on the Ly$\alpha$ images, a different method was used for each object depending on the information available in the data.
For MRC 2025-218, we placed  the radio core  at the position of the
continuum centroid in the VIMOS data. Since this is a broad line object (\S3.2), it is
reasonable to expect that this   marks the location of the active nucleus (AGN). This is further supported by the
fact that the UV continuum has an spatially unresolved component (\S3.2).

For MRC 1558-003, we positioned the radio core at the spaxel with the 
maximum Ly$\alpha$ flux, which is expected to be shifted by a few tenths of
a spaxel relative to the continuum centroid (Humphrey et al. \citeyear{hum07}). This method  would be incorrect if the AGN
is spatially shifted relative to the continuum  and/or line centroids. However,
 we do not expect this to be the case 
since this is a broad line object and moreover, it shows little evidence for Ly$\alpha$ absorption (see below).  The  1$\sigma$ uncertainty
 in the Ly$\alpha$-radio
registration is $\sim$0.3" in both cases.

For MRC 0140-257, we assumed that the radio core is placed between the two Ly$\alpha$ spatial
components, although it is not necessarily the case. This is the
 main source of uncertainty since
 the radio core could be shifted by 2$\arcsec$ if it is associated with one of the two Ly$\alpha$ blobs (\S4.3). The impact of  this assumption on the
interpretation of our results will be discussed when relevant.

\subsection{Vimos sensitivity}

The main  scientific goal of our VIMOS-VLT observational program
is to study the  properties
of the giant  nebulae associated with a sample of HzRG. By isolating
spectroscopically and/or spatially  the emission from the perturbed   and  the quiescent gaseous components
(e.g. Villar-Mart\'\i n et al. \citeyear{vm03}), one of our priorities
is to
characterize
the morphological, kinematic and ionization properties of the quiescent
gas. In this way we can study the gas without the strong distortions  that 
the radio structures can imprint  on its observed properties.
 Such study has the potential to provide critical information on the star formation
and chemical enrichment histories as well as the galaxy formation process 
(Villar-Mart\'\i n et al. \citeyear{vm03}, Humphrey et al. \citeyear{hum06}, \citeyear{hum07}). 

In order to achieve
these goals, it is critical to    detect the main UV lines 
(very especially Lya and HeII)  with high  signal/noise  in the faint, outer regions 
of the objects, which are usually characterized by quiescent kinematics.
Moreover, this  would allow to 
study in detail
the high surface brightness regions (at least several times brighter).
Although usually distorted by the radio structures,
it should  be possible to isolate spectroscopically the emission from the quiescent
and the perturbed gas using especially the non resonant HeII line, as Villar-Mart\'\i n et al. (\citeyear{vm03}).

The faintest regions of our interest   have 
often  Ly$\alpha$ surface brightness 
levels which are, within the errors, consistent with or even below 
 the detection limit
or our Vimos data (3$\sigma\sim$10$^{-17}$ erg s$^{-1}$ cm$^{-2}$ arcsec$^{-2}$)
 for
8-10 hours exposure time. For comparison,
 van Ojik et al. (\citeyear{ojik96}) detected in 4 hours  the giant, quiescent, low surface
brightness nebula ($\sim$l0$^{-17}$ erg s$^{-1}$ cm$^{-2}$ arcsec$^{-2}$)  associated with a radio galaxy at $z=$3.6 
using
long slit  spectroscopic data obtained with EMMI (ESO Multi-Mode Instrument) on the 3.5m NTT telescope
(La Silla). With a 2.5$\arcsec$ wide slit, the spectral resolution was
 of 2.8 \AA, not very different to that of our data. 
 In the VIMOS data, not even Ly$\alpha$ could  be definitively detected
 from the faintest regions of our interest, which at least in MRC 1558-003 and MRC 2025-218 we knew exist (\S3).  HeII and CIV are detected only from the highest surface brightness regions, but with not enough signal/noise ratio to perform an adequate kinematic and 
ionization  analysis in two spatial dimensions.

Thus, the investigation of the properties of the quiescent gas has been possible
only for those objects where there is no signature of interactions
between the radio structures and the gas (MRC 1558-003 and MRC 0140-257,
see \S4) or objects with no spectroscopic blend between the perturbed
and the quiescent nebula (MRC 2104-242, Villar-Mart\'\i n et al. \citeyear{vm06}). Moreover, in all three cases, Ly$\alpha$ is not
heavily absorbed.
For objecs where the emissions from
the perturbed and the quiescent gas are blended and/or Ly$\alpha$ is absorbed (i.e. an important fraction of 
HzRG),
a more sensitive instrument/telescope combination is needed.

\section{Previous results}

The three radio galaxies discussed in this paper, MRC 1558-003,
MRC 2025-218, MRC 0140-257, 
 belong to the Molonglo
Catalogue of high redshift radio galaxies (Large et al. \citeyear{lar81}, McCarthy et al. \citeyear{mac90b}).
The objects were  selected based on previous  evidence for strong Ly$\alpha$ emission. The two first objects were previously known
to be associated with giant ($>$60 kpc) Ly$\alpha$ nebulae (see below). 

\subsection{MRC 1558-003 (z=2.53)}

This object is associated with  
radio structures which extend for  $\sim$9'' or 84 kpc 
(Pentericci et al. \citeyear{pente00b}) along  a position angle ($PA$) of 75$^o$.

Previous long-slit spectroscopic studies have shown that Ly$\alpha$ is extended
for at least $\sim$14$\arcsec$ ($\sim$130 kpc) along the slit with $PA=$72$^o$,  well beyond the radio structures (e.g. Villar-Mart\'\i n et al. \citeyear{vm03}).  
 CIV and HeII are extended across $\sim$6'', while NV, which
is fainter, has a  more
compact appearance. The optical continuum is also extended.

 ISAAC near infrared spectroscopy revealed 
very broad H$\alpha$ (FWHM$\sim$11700$\pm$900 km s$^{-1}$, Humphrey\citeyear{hum04}; Humphrey
et al. 2007, in prep.), evidence for an obscured broad line region
(BLR).
According to the
unified scheme of quasars and radio galaxies (Barthel \citeyear{bar89}), 
the detection of BLR emission implies that the  ionization
cones axis is at a small angle to line of sight ($<$45$\degr$).
This is further suggested by the detection of a clear one-sided radio jet
 (see Fig.~1).

High resolution optical spectroscopy (FWHM$\sim$1.7 \AA ) showed no 
 absorption features in the Ly$\alpha$ spectral profile integrated along $PA$=85$^o$
(van Ojik et al. \citeyear{ojik97}). Absorption features were not found either
along     $PA$=75$^o$ by Villar-Mart\'\i n et al. (\citeyear{vm03}), 
although the spectral resolution was
rather low in this case (FWHM$\sim$11 \AA). The small impact of
absorption effects is also suggested by the large Ly$\alpha$/HeII values
measured across the nebula (Humphrey \citeyear{hum04}, 
Villar-Mart\'\i n et al. \citeyear{vm07})

For this object no HST or emission line images are available in the literature.

\subsection{MRC 2025-218 (z=2.63)}

This radio galaxy is associated with 
a small radio source ($\sim$5'' or 46 kpc, Carilli et al. \citeyear{car97}).

There is strong evidence for an obscured  BLR in 
MRC 2025-218. The morphology of the object at
  optical  and  near infrared wavelengths  is dominated by a point source (Pentericci et al. 1999, 2001; McCarthy, Person \& West 1992). 
 Broad CIV  and broad H$\alpha$  (FWHM$\sim$6000 km s$^{-1}$ ) revealed
by optical (Villar-Mart\'\i n et al.  \citeyear{vm99}) and
near infrared (Larkin et al. \citeyear{larkin00};
Humphrey\citeyear{hum04}; Humphrey et al. 2007, in prep.)  spectra  confirm this interpretation. The UV rest frame continuum emission shows also an extended
($\sim$6'') 
diffuse component, which 
 is well aligned with the radio axis  (Pentericci et al. 1999).

The Ly$\alpha$ emission is distributed bimodally (McCarthy et al. \citeyear{mac90b}) north-south
with two peaks that correspond roughly to the two radio lobes.
The authors measured a total extension of $\sim$4'' or 37 kpc. Keck
long-slit optical spectroscopy (Humphrey\citeyear{hum04}) 
reveals that Ly$\alpha$ is extended up to $\sim$9'' or 83 kpc. 
CIV, HeII and CIII] 
are also spatially extended. CIV and CIII]  each have a spatially
unresolved component centered at the continuum centroid and also an underlying extended component spanning $\sim$5'' along the radio axis. 
NV is spatially unresolved.

\cite{vm99}  detected absorption in the spectrum
of MRC 2025-218 for CIV, CII$\lambda$1335, SiIV$\lambda\lambda$1393.8,1402.8 and, maybe, OI$\lambda$1302.2+ SiII$\lambda$1402.8. The steep Ly$\alpha$ profile
on the blue side of the line was also proposed to be a signature of absorption.

\subsection{MRC 0140-257 (z=2.64)}

This object is associated with  a small 
double radio source (4.2$\arcsec$ or 39 kpc, Carilli et al. \citeyear{car97}).
In the optical, it appears as a faint galaxy  spatially extended and 
 aligned with the radio source (McCarthy, Persson \& West 1992).

 Ly$\alpha$ and H$\alpha$ are the two lines 
detected in previous spectroscopic studies (McCarthy et al. \citeyear{mac91a}; Eales \& Rawlings \citeyear{eales96}).  It has not been reported whether
the lines are spatially extended.

The F160W filter NICMOS-HST image (rest frame spectral window 3850-4950 \AA, Pentericci et al. \citeyear{pente01}) 
shows two peaks of emission with almost the
same flux, closely aligned with the radio axis. 
The authors suggest that this morphology could be due to a dust lane. There
are a few fainter clumps within 2-3''.
 
\cite{eales96} report Ly$\alpha$/H$\alpha$=0.31, which is $\sim$30 times
lower than standard case B photoionization model predictions. 
As the authors explain, this value is very uncertain, since they had to assume a fixed value
of the [NII]/H$\alpha$ ratio (0.5) and no aperture corrections to the line fluxes
were attempted. In spite of these uncertainties, the very large 
discrepancy with the model predictions suggest that  Ly$\alpha$ absorption/extinction is present.

For this object no emission line images are available in the literature.

\section{Results}

We present below the most relevant results obtained for the three radio
galaxies investigated here.

\subsection{MRC 1558-003 (z=2.53)}

\centerline{\it The  Ly$\alpha$ nebula}

\vspace{0.2cm}

The VIMOS Ly$\alpha$+continuum image of MRC 1558-003 was created by adding 
the monochromatic images of the object within the [4285-4315] \AA\ range.
A continuum image was subtracted to keep the line emission only. 
This  was created by combining
the images extracted from two adjacent spectral windows at both sides
of Ly$\alpha$, with the same
spectral width as the line+continuum image. The resulting Ly$\alpha$ image
is shown in Fig.~1 (see also Fig.~2), with the 8.2GHz radio contours
overlaid (Pentericci et al. \citeyear{pente00b}).

The nebula extends across $\sim$9''$\times$7.5'' or 84$\times$70 kpc$^2$
(but see \S3.1).
 It is characterized by a strongly peaked rounded morphology. Emission
from the  nebula  is detected only within the radio structures.
 The faintest
Ly$\alpha$ emission we detect on the VIMOS image 
has surface brightness $\sim$10$^{-17}$ erg s$^{-1}$ cm$^{-2}$ arcsec$^{-2}$
(3$\sigma$ values).

The angle between the nebular axis and the radio structures was measured
using the position angle on the sky of the
longest dimension of the Ly$\alpha$ nebula and the position angle of 
 the line between the two brightest radio hot spots. The nebula is
 misaligned by $\sim$30$\pm$5$^o$ relative to the radio axis.

As we found for MRC 2104-242 (Villar-Mart\'\i n et al. \citeyear{vm06}), Ly$\alpha$
emission is detected outside any plausible ionization cones with opening
angle $\le$90$^o$ (Barthel \citeyear{bar89}).
Seeing effects
are not likely to be responsible for this Ly$\alpha$ emission since the observations
of this object were carried out  under seeing conditions
of FWHM$<$ 1.5$\arcsec$. Since
this is a broad line object, it is possible 
that the broad rounded  morphology  of the Ly$\alpha$ nebula
is due to orientation effects (see \S5.3). If the ionization cone axis lies close to the line
of sight, a broader, rounder morphology is expected.

The total Ly$\alpha$ flux integrated over the nebula is 4.8$\times$10$^{-15}$ 
erg s$^{-1}$ cm$^{-2}$ corresponding to a luminosity of 3.4$\times$10$^{44}$
erg s$^{-1}$.

\begin{figure}
\includegraphics{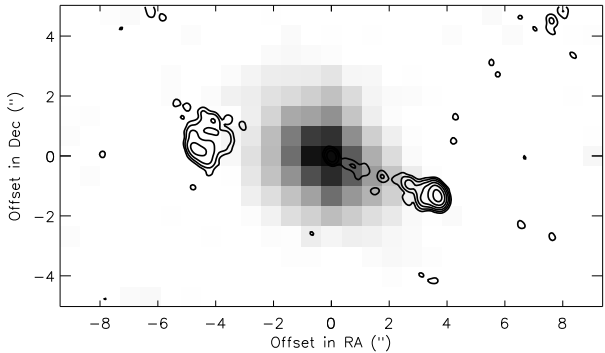}
\vspace{2.0in}
\caption{Ly$\alpha$ nebula (grey scale) associated with MRC 1558-003 with the  8.2GHz radio contours 
 overlaid. The maximum extension of the nebula, as measured from the VIMOS data, is
$\sim$9''$\times$7.5''   (84$\times$70 kpc$^2$).
 The nebula is misaligned by $\sim$30$\pm$5$^o$ relatively to the
radio structures. North-up; East-left.}
\end{figure}

\vspace{0.2cm}
\centerline{\it CIV and HeII morphologies}

\vspace{0.2cm}

The CIV and HeII images were created by adding the monochromatic
images of the object within the spectral windows  
[5455-5485] \AA\ and [5780-5800] \AA\
 respectively. Continuum images adjacent in wavelength to
each line and of the same spectral width
were subtracted to keep the line emission only. 

The resulting images are shown in Fig.~2. The Ly$\alpha$ contours have
been overplotted on the top panels for comparison.
The CIV  line is  spatially extended 
 with a maximum projected
size of  $\sim$7.5$\arcsec$. The outer, low surface brightness regions
extend in the same direction as Ly$\alpha$. 
HeII, which is fainter, appears more rounded and compact, although
it is spatially resolved with a maximum
extension of $\sim$3$\arcsec$. 

The inner ($\sim$2$\arcsec$) regions of the CIV nebula seem
to be extended along an axis (roughly E-W) which is shifted by $\sim$15$\degr$ anti-clockwise relative to the axis defined by the outer regions and 
aligned within a few degrees with the radio structures. The HeII emission seems to be extended also  in this direction. There is some hint that this could
also be the case for the inner Ly$\alpha$ nebula. However, higher spatial resolution data would be necessary
to investigate whether there is a real rotation of the nebular
axis as we move outwards, which could be a consequence of  an inhomogeneous distribution of
material.

Continuum is also detected, although the image is too noisy 
to characterize its morphology and determine its spatial centroid.

\begin{figure}
\includegraphics{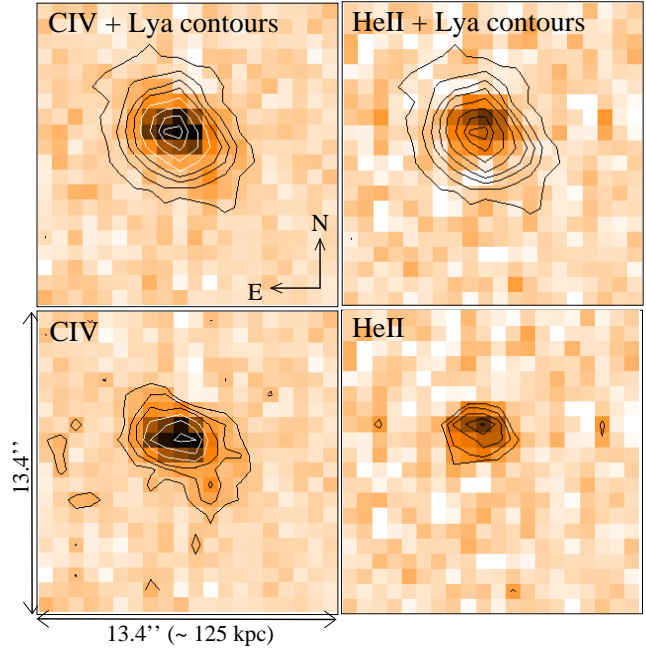}
\vspace{3.5in}
\caption{MRC 1558-003: Comparison between the CIV, HeII and Ly$\alpha$ spatial distributions
(see electronic manuscript for the colour version fo the figures).
The underlying continuum has been subtracted from all images.
Top panels: The Ly$\alpha$ morphology is shown with contours overlaid on
the CIV (top-left) and HeII (top-right) images (colour scale). The bottom panels show
the CIV and HeII morphologies.
Ly$\alpha$ contours: (0.15, 0.6, 1.0, 1.6, 2.0, 3.2, 4.4, 4.4, 5.0)$\times$10$^{-16}$ erg s$^{-1}$
cm$^{-2}$ arcsec$^{-2}$; CIV contours: (1.8, 2.5, 4.1, 5.5, 7.0, 8.3)$\times$10$^{-17}$ erg s$^{-1}$  cm$^{-2}$ arcsec$^{-2}$; HeII contours: (1.8, 2.4, 3.2, 3.9, 4.6, 5.3)$\times$10$^{-17}$ erg s$^{-1}$
cm$^{-2}$ arcsec$^{-2}$.}
\end{figure}
\vspace{0.2cm}
\vspace{0.2cm}
\centerline{\it Spatially extended emission line spectroscopy}

\vspace{0.2cm}

We show in Fig.~3  the spatial maps of the FWHM (corrected for instrumental
profile) and velocity shift $V_{offset}$ of the Ly$\alpha$ line. These values have
been measured fitting Gaussian profiles spaxel to spaxel. $V_{offset}$ has been computed  relative to the
Ly$\alpha$ emission at the spatial line centroid. The errors on the FWHM and $V_{offset}$ 
are estimated to be, in general, 
$<$100 km s$^{-1}$ and $<$40 km s$^{-1}$ respectively.

Although the
line profile is asymmetric at some spatial positions, this method
allows a first order study of the spatial variation of the Ly$\alpha$
spectral profile across the nebula.
The line spectral profile shows strong spatial variations, as is obvious from Fig.~3.

 The first striking characteristic is that Ly$\alpha$ presents an increasing redshift towards the nebular centroid, 
with a maximum shift in velocity of $\sim$400 km s$^{-1}$ at this position relative to the outer regions.

 The FWHM varies between $\sim$450 and
1350 km s$^{-1}$  across the nebula. Three distinct regions ($A$, $B$ and $C$
in Fig.~3) can be isolated in the FWHM map according to
the line width. The maximum  values are measured in region $B$ (in the range
1050-1340 km s$^{-1}$), which runs approximately along the radio axis
and contains the nebular centroid. The HeII FWHM  measured
from the integrated spectrum is 600$\pm$100  km s$^{-1}$.
Region $A$ shows narrower Ly$\alpha$, although the line is still quite
broad (FWHM in the range
700-1030 km s$^{-1}$) while the  HeII FWHM is  650$\pm$200 km s$^{-1}$ in the integrated spectrum. At some positions
  HeII  is as narrow as 450$\pm$50 km s$^{-1}$ (consistent with
Villar-Mart\'\i n et al. \citeyear{vm03}). Ly$\alpha$ trends to be notably
broader than HeII (this was also found by Villar-Mart\'\i n et al.
\citeyear{vm03} along the radio axis).

Region $C$ is clearly different both in line width (FWHM(Ly$\alpha$)=650$\pm$50 
km s$^{-1}$, Fig.~4, bottom) and velocity shift. This region   shows the largest blueshift relative to the Ly$\alpha$
centroid (350$\pm$20 km s$^{-1}$ for the integrated spectrum). 
  CIV is
detected in region $C$, although noisy, and it has  FWHM=600$\pm$100 km s$^{-1}$.

\begin{figure}
\includegraphics{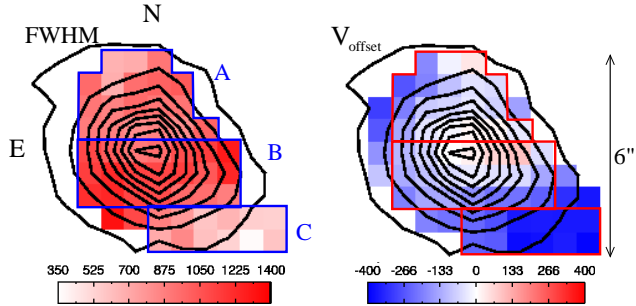}
\vspace{2.0in}
\caption{MRC 1558-003: 2-dim Ly$\alpha$ spectral properties  (see electronic
version for colour figures). The FWHM (left) 
(corrected for instrumental broadening) and the velocity shift (right) relative to
the line emission at the Ly$\alpha$ spatial centroid are shown. Values
in km s$^{-1}$.
 Ly$\alpha$
flux contours are overplotted. Regions  $A$, $B$ and $C$ (see text) are identified.
(Notice that the spaxels  on which the FWHM and $V_{offset}$ are determined 
are not the same  and depend on the signal to noise ratio).}
\end{figure}

\vspace{0.2cm}

\centerline{\it  Ly$\alpha$ absorption}

\vspace{0.2cm}

 We do not find definitive
Ly$\alpha$ absorption features in the VIMOS data across the nebula.
This is consistent with previous studies (\S3.1).

\vspace{0.2cm}

\subsection{MRC 2025-218(z=2.63)}

\centerline{\it The Ly$\alpha$ nebula}

\vspace{0.2cm}

The VIMOS Ly$\alpha$+continuum image of MRC 2025-218 was created by adding 
the monochromatic images of the object within the [4385-4445] \AA\ range.
A continuum image extracted from an adjacent line-free region
of the same spectral width was subtracted to keep the line emission only. 
 The resulting Ly$\alpha$ image is shown in Fig.~4, with the VLA 8.2GHz radio contours
overlaid (Carilli et al. \citeyear{car97}; see also Fig.~5). 

\begin{figure}
\includegraphics{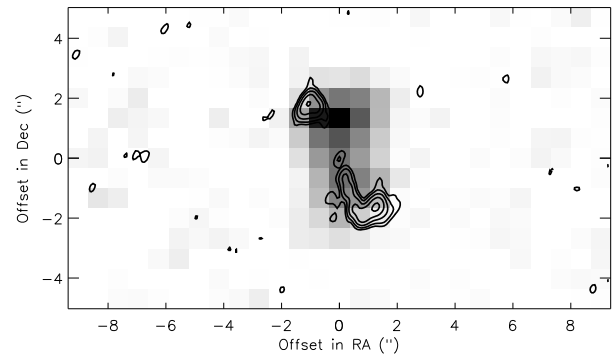}
\vspace{2.1in}
\caption{Ly$\alpha$ nebula (grey scale) associated with MRC 2025-218 with
 8.2GHz radio
contours  overlaid. The maximum extension of the nebula as measured
from the VIMOS data is
$\sim$6'' or  55  kpc. The nebula is similar in size to the
radio source and is closely aligned with it (but see text). North-up; East-left.}
\end{figure}

 The Ly$\alpha$ maximum extension
is $\sim$6.0'' or $\sim$55 kpc  (but see \S3.2), very similar to the radio source size
.  The faintest
Ly$\alpha$ emission we detect on the VIMOS image 
has surface brightness $\sim$1.5$\times$10$^{-17}$ erg s$^{-1}$ cm$^{-2}$ arcsec$^{-2}$ (3$\sigma$ values). The maximum extension in the direction perpendicular to the radio structures
as measured from the VIMOS data is $\sim$4$\arcsec$.
 As  \cite{mac90b} already pointed out,
the nebula is aligned
with the radio structures within a few degrees. The bimodal distribution described in \S3.2 is clearly
seen.

The total Ly$\alpha$ flux integrated over the nebula is 2.1$\times$10$^{-15}$ 
 erg s$^{-1}$ cm$^{-2}$ corresponding to a luminosity of 1.7$\times$10$^{44}$
erg s$^{-1}$.

\vspace{0.2cm}

\centerline{\it CIV and continuum morphologies}

\vspace{0.2cm}

The CIV image (HeII is too faint) 
was created by adding the monochromatic
images of the object within the spectral window [5610-5650] \AA\  and
subtracting the adjacent continuum.  The continuum image of the object was
created by integrating across the 
[5670-5850] \AA\ spectral window.

The resulting images are shown in Fig.5. The Ly$\alpha$ contours have
been overplotted (top panels)  for comparison.

CIV extends  for $\sim$4$\arcsec$ in the same direction 
as Ly$\alpha$ (N-S). The bimodal distribution shown by Ly$\alpha$ is
distinguished in the CIV image. 
Interestingly, the continuum is spatially unresolved in 
this direction, but is barely resolved in the E-W direction, with 
a FWHM of $\sim$1.1$\pm$0.1$\arcsec$, compared with FWHM=1.5$\pm$0.2$\arcsec$
for the star in the field. The continuum centroid is located between
the two Ly$\alpha$ spatial components, i.e., where this line presents
a minimum.

\begin{figure}
\includegraphics{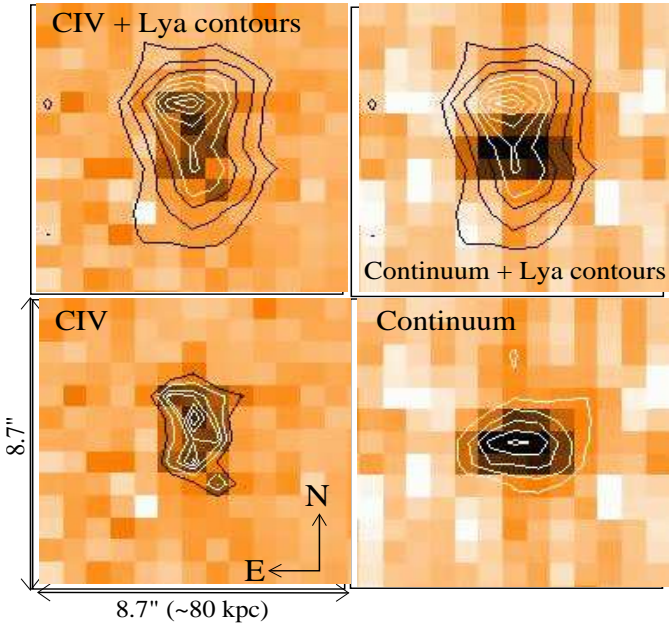}
\vspace{3.4in}
\caption{MRC 2025-218: Comparison between the CIV, continuum 
and Ly$\alpha$ spatial distributions.
Top panels: The Ly$\alpha$ morphology is shown with contours overlaid on
the CIV (continuum
subtracted, left) and continuum (right) images. Bottom panels: CIV  (left) and continuum (right) images. 
CIV is extended in the same direction as Ly$\alpha$.  The continuum 
centroid is located between the two Ly$\alpha$ spatial components, where
the line flux presents a minimum. Ly$\alpha$ contours: (0.2, 0.5, 1.1, 1.4, 1.7, 2.1,  2.7, 3.0)$\times$10$^{-16}$ erg s$^{-1}$ cm$^{-2}$ arcsec$^{-2}$.
 CIV contours: (1.4, 2.0, 2.5, 2.8, 3.2, 3.5)$\times$10$^{-17}$ erg s$^{-1}$ 
cm$^{-2}$ arcsec$^{-2}$. Continuum contours: (1.7, 3.2, 4.5, 6.0, 7.5)$\times$10$^{-16}$ erg s$^{-1}$ 
cm$^{-2}$ arcsec$^{-2}$.}
\end{figure}

\vspace{0.2cm}

\centerline{\it Spatially extended emission line spectroscopy}

\vspace{0.2cm}

We show in Fig.~6 the spatial maps of the FWHM  and velocity shift $V_{offset}$ of the Ly$\alpha$ line. As before, 
these values have
been measured by fitting Gaussian profiles to the line spaxel by spaxel.
 $V_{offset}$ has been computed  relative to the
Ly$\alpha$ emission at the position of the continuum centroid.  The errors on the FWHM and $V_{offset}$ 
are estimated to be, in general, 
$<$100 km s$^{-1}$ and $<$35 km s$^{-1}$ respectively.

\begin{figure}
\includegraphics{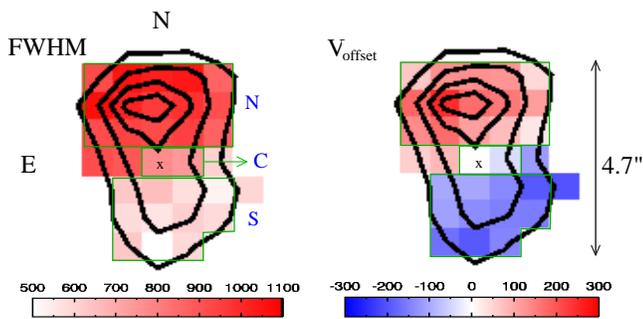}
\vspace{2.0in}
\caption{MRC 2025-218: 2-dim Ly$\alpha$ spectral properties. The FWHM (left) 
(corrected for instrumental broadening) and the velocity shift (right) relative to
the line emission at the continuum spatial centroid are shown.  Values
in km s$^{-1}$.The location of the continuum centroid is shown with
an 'x'. Ly$\alpha$
flux contours are overplotted. Regions $N$, $S$ and $C$ (continuum centroid)
discussed in the text are overplotted.
}
\end{figure}

This analysis reveals two clearly distinct regions (north and south), which are different both
in FWHM  and  $V_{offset}$. 
These are  coincident with the two Ly$\alpha$ spatial 
components discussed above.  The continuum
centroid (located between both regions) shows intermediate FWHM 
 values. 
The Ly$\alpha$ spectra extracted from the apertures $N$ and $S$ (north
and south regions, Fig.~6) 
and the
region in between (2 spaxels, see aperture $C$ in 
Fig.~6) are shown in Fig. 7 (left panels). Both regions emit also CIV and HeII
(Fig.7, top-right panel). 
The  CIV-HeII spectra
 (Fig.~7, top right panel) have been extracted  
using ony the highest flux spaxels  in
the N and S regions (5 and 4 spaxels respectively) in order to maximize
 the signal to noise ratio for the lines. 
The differences in FWHM and $V_{offset}$ seen in Fig.~6
are obvious also here. 

The $S$ region shows the narrowest Ly$\alpha$ spectral profiles
(FWHM in the range 500-650 km s$^{-1}$) and the gas is blueshifted relative
to the continuum centroid
(values in the range $\sim$-100 to -200 km s$^{-1}$). CIV (which is a doublet and, therefore, expected
to be intrinsically broader) and HeII are also rather narrow, with FWHM=650$\pm$30 and 500$\pm$30 km s$^{-1}$ respectively, as measured from the spatially
integrated spectrum.  

The $N$ region is characterized by broader emission lines 
(FWHM in the range $\sim$850-1070 km s$^{-1}$ for Ly$\alpha$) and the gas is redshifted relative to the continuum centroid ($V_{offset}$ in the range $\sim$ +50 to +250
 km s$^{-1}$). The CIV FWHM is 1220$\pm$40 km s$^{-1}$. HeII is too noisy in the VIMOS
spectrum to measure its FWHM. However,  Keck long-slit spectroscopy
 (Humphrey \citeyear{hum04}) of this region implies
 FWHM=1100$\pm$100 km s$^{-1}$. Therefore, the three lines in the N region
are very broad compared with the S region.

Although absorption plays an important role in the Ly$\alpha$ spectral differences across the nebula (see below), kinematics is also clearly having an effect, since
the CIV and HeII lines show marked differences between the N and S regions.

The Ly$\alpha$ and CIV spectra at the position of the continuum centroid are shown in Fig.~7 (bottom panels). Very broad wings (presumably from the BLR, see \S3.2) 
are seen in the line profiles of both Ly$\alpha$ and CIV.

\begin{figure}
\includegraphics{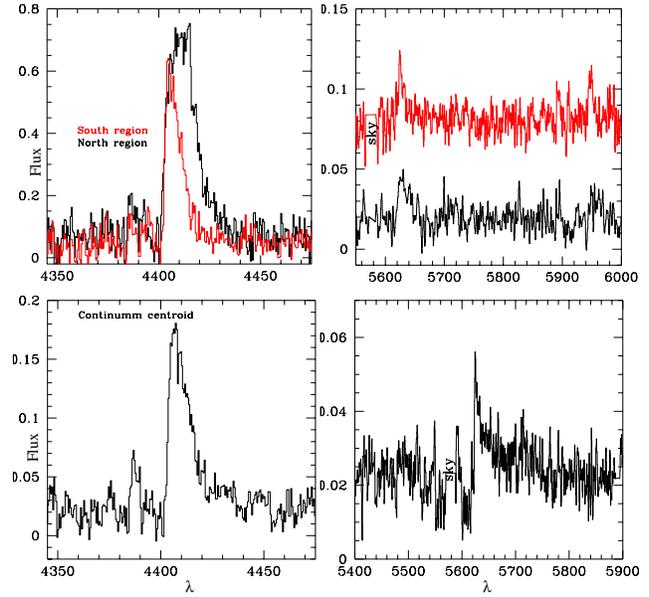}
\vspace{3.2in}
\caption{MRC 2025-218. Top: Comparison between  the  Ly$\alpha$ (left) and CIV-HeII (right) spectra
 from the
northern and southern regions (see text). The southern CIV-HeII spectrum
has been shifted  artificially in the flux scale for clarity. The two lines are detected in both regions. 
Bottom: Ly$\alpha$ (left) and CIV (right) spectra at the position of the continuum
centroid.
Notice the very broad wings of both lines, signature of the broad line region,
as well as
the absorption features. In spite of the
clear difference in FWHM, the shift in $\lambda$ of the Ly$\alpha$ line
  and the spatial separation between the two regions, 
notice the exact coincidence in wavelength
of the sharp edge in the blue wing of the line for the three spectra. 
Fluxes in units of
10$^{-16}$ erg s$^{-1}$ cm$^{-2}$ \AA$^{-1}$.  }
\end{figure}

\vspace{0.2cm}

\centerline{\it Ly$\alpha$ and CIV absorption}

\vspace{0.2cm}

The Ly$\alpha$ and CIV lines are clearly absorbed  in this object
 (Fig.~7),
as was already discussed by other authors (Villar-Mart\'\i n et al. \citeyear{vm99}). 
Since this paper is mostly focused on the properties of the 
emission line nebulae, we will present here a general description
 and  defer a  more detailed
 analysis of the absorbers for another publication 
(Humphrey et al. 2007, in prep).

Some peculiar properties of the Ly$\alpha$ spectral  profile 
are a consequence of  absorption rather
than kinematics: the   multiple peaks, the fact that the flux
drops below the continuum level at some wavelengths, the sharp,
almost vertical 
edge of the blue wing of the line, and the identical wavelength 
over the nebula at which this
sharp edge is measured (see below)  are most naturally explained
by absorption.   Several absorption features are detected in both
Ly$\alpha$ and CIV (Fig.~7).

Ly$\alpha$ absorption is detected in more than 30 spaxels. 
We set a lower limit
to the size of the main absorber of $\sim$4.7$\arcsec$$\times$3.5$\arcsec$
or $\sim$43$\times$32 kpc$^2$.
It is possible that this absorber covers the Ly$\alpha$ nebula completely. 
Since absorption is detected in CIV as well, this implies that
the absorbing gas is highly ionized.

A striking characteristic  is 
   that the sharp edge of the Ly$\alpha$ blue wing (see Fig.~7) 
happens at almost exactly the same
wavelength (4403.3$\pm$0.5 \AA) in all spaxels where we have
been able to measure it ($\sim$20), revealing very little kinematic
structure of the main absorber along the line of sight and accross
its whole spatial extension.

\subsection{MRC 0140-257 (z=2.64)}

\centerline{\it  The  Ly$\alpha$ nebula}
\vspace{0.2cm}

The VIMOS Ly$\alpha$+continuum image of MRC 0140-257 was created by adding 
the monochromatic images of the object within the [4410-4430] \AA\ range.
An adjacent continuum image was subtracted to keep the line emission only. 
 The resulting Ly$\alpha$ image
is shown in Fig.~8 with the VLA 8.2GHz  radio contours    overlaid (Carilli et al. \citeyear{car97}). 

Two main spatial components ($A1$ and $A2$ in the figure) 
are identified in the image, which are aligned with
the radio structures within a few degrees. We have assumed that the radio core is located between
 $A1$ and $A2$ (but see \S2).
 The  Ly$\alpha$ spatial
centroids of $A1$ and $A2$
are separated by $\sim$3.5$\arcsec$ or 32 kpc. 
The maximum extension of the  nebula ($A1$ and
$A2$) is
9$\arcsec$  or  83 kpc. 
 It, therefore, extends well beyond the radio
structures. This conclusion is independent of the location of
the radio core along the $A1$-$A2$ line (\S2). The faintest
Ly$\alpha$ emission we detect on the VIMOS data 
has surface brightness $\sim$10$^{-17}$ erg s$^{-1}$ cm$^{-2}$ arcsec$^{-2}$ (3$\sigma$ values).

\begin{figure}
\includegraphics{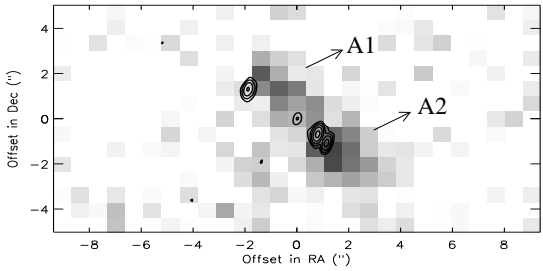}
\vspace{1.8in}
\caption{Ly$\alpha$ nebula associated with MRC 0140-257 with 8.2GHz radio contours
overlaid. 
Two main Ly$\alpha$ spatial components ($A1$ and $A2$) are identified in the image
very closely aligned with the radio structures.
The  size of the nebula is $\sim$9$\arcsec$  or  83 kpc. North-up; East-left.}
\end{figure}

The total Ly$\alpha$ flux integrated over the nebula is 3.6$\times$10$^{-16}$ 
 erg s$^{-1}$ cm$^{-2}$, which, taking aperture corrections into account,
is consistent with  \cite{mac91a}. It 
corresponds to a luminosity of 2.9$\times$10$^{43}$
erg s$^{-1}$.

\vspace{0.2cm}

\centerline{\it Continuum morphology}

\vspace{0.2cm}

A continuum image was created by collapsing the data
cube across the continuum 
spectral windows [5270-5550] \AA\ and [5640-5800] \AA, i.e.  rest frame range $\sim$[1450-1530] \AA\ and [1555-1600] \AA\ respectively. For comparison, the optical image of McCarthy, Persson \& West (\citeyear{mac92}) 
was obtained using a Gunn-Thuan filter, covering the rest frame range [1675-1920].\footnote{The authors used
a 2.5m telescope and an exposure time of 2700 sec to obtain this image.}.
 We detect a faint source, which overlaps
partially with 
component $A1$ (Fig.9). Due to the large separation in wavelength between
the continuum and Ly$\alpha$ images, we cannot discard that 
the spatial shift between the continuum centroid and $A1$ 
is partially artificial (see \S2). 

Although the source is apparently extended in the N-S direction, this
is dubious. The faintest
structures to the north are consistent within the errors with the image detection limit. On the other hand, 
there are noise residuals at similar flux levels at different positions accross the VIMOS field of view. 

\begin{figure}
\includegraphics{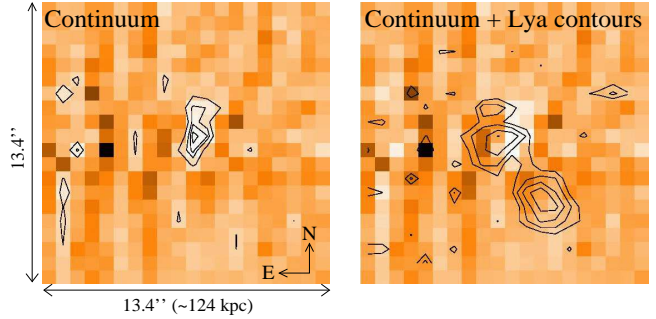}
\vspace{1.8in}
\caption{MRC 0140-257: Comparison between the  continuum 
and Ly$\alpha$ spatial distributions.
Left: continuum  images. Right: With Ly$\alpha$ contours overlaid.
 Ly$\alpha$ contours: (0.2, 0.4, 0.6, 1.0)$\times$10$^{-16}$ erg s$^{-1}$ cm$^{-2}$ arcsec$^{-2}$. Continuum contours: (2.1,3.5,4.2,4.9)$\times$10$^{-16}$ erg s$^{-1}$ 
cm$^{-2}$ arcsec$^{-2}$.}
\end{figure}
\vspace{0.2cm}

\centerline{\it Spatially extended emission line spectroscopy}

\vspace{0.2cm}

We show in Fig.~10 the spatial maps of the FWHM and $V_{offset}$ of
the Ly$\alpha$ line. $V_{offset}$  has been measured relative to the
Ly$\alpha$ emission between $A1$ and $A2$. The erros  on $V_{offset}$
are estimated to be $<$40 km s$^{-1}$. The errors on the FWHM 
are in the range 60-100 km s$^{-1}$. The main reason for these
large relative errors is the uncertainty
on the continuum level,  due to  the noise (this object is 
fainter) and the presence
of an underlying broad component (see below).
The line is  narrow,
with FWHM$\la$500 km s$^{-1}$ accross the whole nebula, compared
with typical values of HzRG.

\begin{figure}
\includegraphics{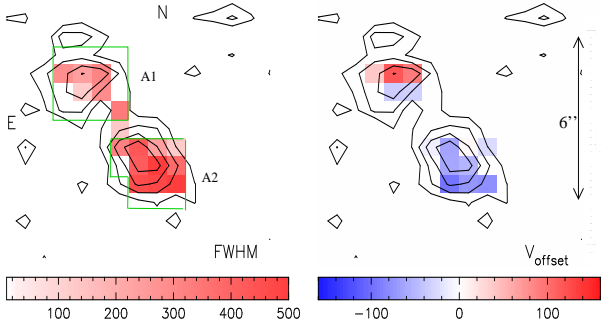}
\vspace{2.0in}
\caption{MRC 0140-257: 2-dim Ly$\alpha$ spectral properties. The FWHM (left) 
(corrected for instrumental broadening) and the velocity shift (right) relative to
the line emission at the intermediate spatial position between $A1$ and
$A2$ are shown.  Values
in km s$^{-1}$.  Only coloured (i.e., not white) spaxels in the FWHM map
 have measured
FWHM values. $V_{offset}$ is shown for the same spaxels, with
white corresponding in this case to 0 velocity. 
Ly$\alpha$
flux contours are overplotted.  The apertures selected to extract the
spectra of components $A1$ and $A2$ (see text) are shown with green lines.}
\end{figure}

\begin{figure}
\includegraphics{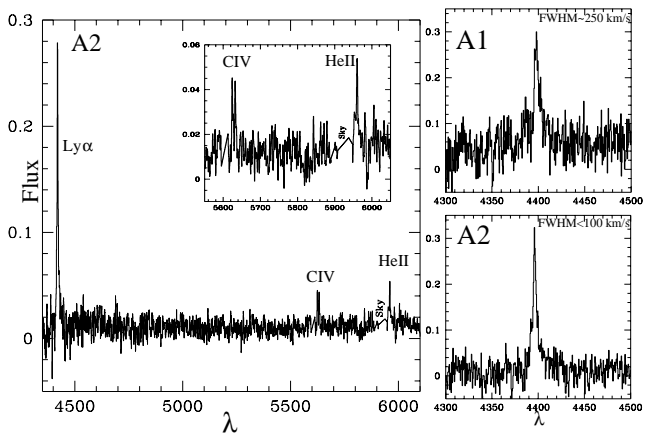}
\vspace{2.6in}
\caption{MRC 0140-257: Spectra of       $A1$ and $A2$. The Ly$\alpha$ spectral region is amplified in the right panels. The complete spectrum is shown
for $A2$ (left panel) to highlight the detection of CIV and HeII (small box).
 All lines are characterized by a very narrow component of FWM$<$300 km s$^{-1}$. The lines are unusually
narrow for a high redshift radio galaxy.
  Notice that the CIV doublet is resolved.
Flux in units of 10$^{-16}$ erg s$^{-1}$ cm$^{-2}$ \AA$^{-1}$.}
\end{figure}

The Ly$\alpha$ spectra integrated over the highest signal to noise spaxels
of  $A1$ (16 spaxels) and $A2$ (14 spaxels, see Fig.~10, green lines)  
are shown in Fig.~11 (right panels).  In both spatial components, 
the Ly$\alpha$ spectral profile is dominated by a strikingly narrow  component,
with FWHM=250$\pm$50 and $\la$120 km s$^{-1}$ for $A1$ and $A2$ respectively.
An underlying broad component seems to be also present in both components. 
\footnote{Using a single Gaussian, as the fits used to produce
Fig.~11 (left) the narrow peak is broadened because of the broad
wings and the derived FWHM have values  of  up to $\sim$500
km s$^{-1}$.}
This is most clearly detected
when both spectra are added. The fit to the line
profile in the coadded spectrum of $A1$ and $A2$ is shown in Fig.~12
together with the individual components isolated in the fit. 
The underlying broad component has FWHM=1200$\pm$200   km s$^{-1}$. 
The velocity shift between  $A1$ and $A2$ is 120$\pm$20 km 
s$^{-1}$, which
is also rather  low compared with typical values in HzRG (e.g.
McCarthy et al. \citeyear{mac96a}).

\begin{figure}
\includegraphics{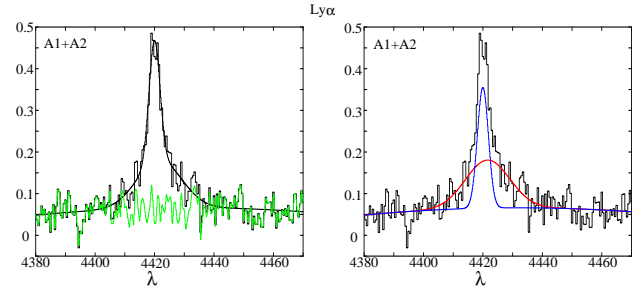}
\vspace{1.8in}
\caption{MRC 0140-257: Ly$\alpha$ spectrum of the coadded spectra of   $A1$ and $A2$. 
The original data are shown with the best fit
(thin solid line) and the residuals (green).  The individual kinematic components (blue and red) are shown
in the right panels.  A 
strikingly narrow 
 component of FWHM=270$\pm$40 km s$^{-1}$ is isolated on top of an underlying
broad components   with FWHM=1200$\pm$200 km s$^{-1}$.
Flux in units of 10$^{-16}$ erg s$^{-1}$ cm$^{-2}$ \AA$^{-1}$.}
\end{figure}

CIV and HeII are detected in $A2$ (Fig.~11, left). Two very narrow
emission lines are identified at the expected wavelengths of the CIV
doublet. We have fitted the lines with no constraints applied.
The resulting doublet consists of two emission lines  separated by 8.0$\pm$1.0 \AA\ (consistent within the errors with the theoretical
9.2 \AA\ at the redshift of the object) and with flux ratio of
$\sim$1.0$\pm$0.2, 
as expected in the optically thin case. Taking the errors into account, both  lines are unresolved (observed
FWHM=2.5$\pm$0.6 and 2.4$\pm$0.7 \AA\ respectively),
with FWHM$\la$100 km s$^{-1}$. This is in excellent agreement with the width of
the Ly$\alpha$ narrow component. HeII is also detected and is similarly
narrow with FWHM=180$\pm$60 km s$^{-1}$.

The small FWHM values of  the CIV and HeII lines     confirm that  Ly$\alpha$ is 
intrinsically very narrow  (rather than absorbed).

\vspace{0.5cm}

\centerline{\it Ly$\alpha$ absorption}

\vspace{0.2cm}

The Ly$\alpha$ spectral profile does not show clear  evidence for  absorption.
This is further supported by the large Ly$\alpha$ ratios. In $A2$
Ly$\alpha$/HeII and Ly$\alpha$/CIV are 14.0$\pm$1.5 and 7.4$\pm$1 respectively.
These values are consistent with standard photoionization model predictions for the measured CIV/HeII=1.8$\pm$0.3
(Villar-Mart\'\i n et al. 2007). 
 Ly$\alpha$/HeII and Ly$\alpha$/CIV  are
 $\ga$ 6 and $\ga$5 respectively for $A1$ and it is not possible to say whether
Ly$\alpha$ is absorbed. The values for the integrated
spectrum are
$\ge$15 and 9$\pm$2 respectively 
which for the measured CIV/HeII$\ga$1.5 do not   imply absorption either.
This is in contradiction with  \cite{eales96} 
 (see \S3.3).

\section{Discussion}

\subsection{Main properties of the Ly$\alpha$ nebulae}

The three radio galaxies investigated here are associated with giant
($>$60 kpc) Ly$\alpha$ nebulae. This was already known for
MRC 1558-003 and MRC 2025-218, but not for MRC 0140-257.
The total Ly$\alpha$ luminosities are $\sim$(0.3-3.4)$\times$10$^{44}$
erg s$^{-1}$, within the range of typical values measured for HzRG.

  The morphologies are varied.  The nebula of one source has a centrally
peaked, rounded appearance (MRC 1558-003). In the other two objects 
it consists of two spatial components. This bimodal morphology could be a consequence  of an obscuring dust structure (e.g. Knopp \& Chambers \citeyear{knop},
Reuland et al. \citeyear{reu03}).
 For MRC 2025-218
this is supported  by the fact that the continuum
centroid coincides with a minimum in the Ly$\alpha$ flux.  \cite{pente01} also proposed the existence of a dust
lane in MRC 0140-257 to explain the near infrared continuum morphology.

 The alignment between the Ly$\alpha$ nebulae and the radio structures is remarkable
in MRC 2025-217 and MRC 0140-257 ($\la$10$^o$).
In MRC 1558-003, the nebula is  misaligned  
by 30$\pm$5$^o$. Similar values have been measured for 
other HzRG radio galaxies (McCarthy, Spinrad \& van Breugel \citeyear{mac95}).  

Based on previous, deeper spectroscopic studies and this work, 
we conclude that, although the high surface brightness emission
tends to be confined within the radio lobes, Ly$\alpha$ emission 
is also detected  beyond the radio structures in the three objects
studied here (see also Villar-Mart\'\i n et al. \citeyear{vm03} for other examples).

 In one case (MRC 1558-003), the new data reveal
 Ly$\alpha$ emission
 outside any plausible ionization cone and far from the radio structures
(see Reuland et al. \citeyear{reu03},  Villar-Mart\'\i n
et al. \citeyear{vm06} for other examples). Projection effects might be responsible,
since this is a broad line object and this is our favoured explanation
(see \S5.3). 
Alternatively, part of the Ly$\alpha$ emission
might be resonantly scattered 
 or powered by a mechanism not related to the active nucleus,
such as young stars (Villar-Mart\'\i n et al. \citeyear{vm07}). Unfortunately,
the data are not deep enough to check whether lines other than Ly$\alpha$ are emitted in these regions. Cooling radiation
(Haiman, Spaans \& Quataert \citeyear{hai00}) is an interesting possibility often discussed in the subject of radio quiet (e.g. Nilsson et al. \citeyear{nil06}) and radio loud
Ly$\alpha$ nebulae. However, the Ly$\alpha$ surface brightness we measure in regions
outside the reach of the ionization cones ($\ga$10$^{-17}$ erg s$^{-1}$
cm$^{-2}$ arcsec$^{-2}$) is too high compared with the model predictions 
(see Villar-Mart\'\i n et al. \citeyear{vm03} for a more detailed discussion on this issue; see also Dijkstra, Haiman \& Spaans  \citeyear{dijk06}).

\subsection{Interactions between the gas and the radio structures}

Interactions between the radio structures (jet and radio lobes) and the ambient gas are known to have
a profound impact on the kinematic properties of the giant nebulae associated with many
HzRG (\S1). 

 For the three objects studied here, this is obvious only in MRC 2025-218, 
in the northern region in particular,  where the emission lines, 
Ly$\alpha$,  HeII and CIV have FWHM$\ga$1000 km s$^{-1}$ (\S4.2). No 
evidence for interactions is found in the southern region, where  the radio structures
present a sharp bend (see Fig.~4). If this has been caused by the collision of the radio structures
with a dense gaseous region\footnote{Since this is a broad line object, projection effects
might exaggerate the angle of the observed bend of the radio structures}, no clear signature has remained in the gas kinematic properties.

Ly$\alpha$ is also very broad across the MRC 1558-003 nebula. However, FWHM$>$1000 km s$^{-1}$ values
 are  measured also in distant regions from the radio structures. Since moreover  HeII is relatively narrow 
(450-650 km s$^{-1}$ across the nebula, see also Villar-Mart\'\i n et al. \citeyear{vm03}), we cannot discard that resonance scattering effects are 
responsible for broadening the Ly$\alpha$ spectral profile.

 In MRC 0140-257  the nebular emission is dominated by  quiescent gas (\S4.3),
i.e., not perturbed by the radio structures.  We cannot tell whether the faint
Ly$\alpha$ underlying broad component (FWHM=1200$\pm$200 km s$^{-1}$) is a consequence of radio/gas interactions
or resonance scattering effects. In fact, the most interesting result for this object
is the  strikingly quiescent kinematics revealed by the bulk of the emission
lines across the nebula  (FWHM$<$300  km s$^{-1}$ and $V_{offset}\sim$120 km s$^{-1}$). 
This characteristic is unique among
 HzRG. 
Similar relaxed kinematics is rather extreme also among 
low $z$ radio galaxies, since only a few show such narrow line widths, usually 
measured in extranuclear regions
 (e.g. Baum, Heckman
\& van Breugel \citeyear{baum90}).

\subsection{Are the nebulae infalling?}

\cite{hum07} proposed that the extended {\it quiescent} ionized
 nebulae  associated with
numerous powerful radio galaxies  at different redshifts are in the process
of infall. These authors searched for correlations between
several radio and optical/UV asymmetries, and found that (i) the quiescent
ionized gas has its highest redshift on the side of the nucleus with the
brightest and more polarized radio hotspot, and that (ii) on the side
where the Lya emission is brightest relative to the other emission lines
and continuum.  They concluded that orientation effects, with the
quiescent gas in infall towards the nucleus, is the most natural scenario
for explaining these correlations.

Our study of MRC 2104-242 based on integral field VIMOS data also suggested that the giant quiescent Ly$\alpha$ nebula associated with this object
could be in the process of infall (Villar-Mart\'\i n et al. \citeyear{vm06}).
However, due to the uncertainty on the 
spatial gas distribution, a rotational pattern could not be discarded. 
It was our later work (Humphrey et al. \citeyear{hum07}) on a larger
sample what allowed us to discard rotation and favour the infall interpretation.

 We investigate next whether the  morphological and kinematic 
 properties of the nebulae
studied here are consistent with infall. 

\vspace{0.2cm}

\centerline{\it MRC 1558-003}

\vspace{0.2cm}

\begin{figure*}
\includegraphics{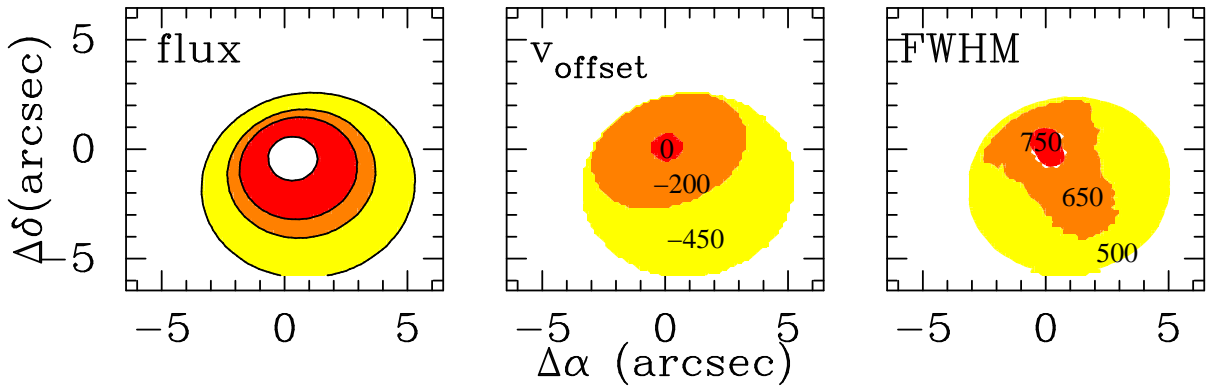}
\vspace{2.1in}
\label{fig:model}
\caption{Observable properties of a model in which an AGN photoionizes a biconical region of the surrounding gas. One cone is pointing almost directly at the observer, while the other points in the opposite direction. The gas is collapsing on to the AGN (for model details, see Appendix~A). If the Ly$\alpha$ flux from the cone that points away from the observer is not detected, then this model reproduces most observational
features (except the Ly$\alpha$ FWHM) remarkably  well: The {\it left panel} shows the surface brightness profile. The contour levels are (0.03,0.01.0.2,0.7) times the maximum surface brightness level. The {\it central panel} shows the velocity shift of the Ly$\alpha$ line with respect to   the centroid (labels denote the off-set in km s$^{-1}$). 
The {\it right panel} shows the FWHM in  km s$^{-1}$ of the Ly$\alpha$ line as a function of position.}
\end{figure*}
According to \cite{hum07}  the giant nebula associated with MRC 1558-003 is infalling towards the center.  
%We discuss here how infall can explain qualitatively the 2-dimensional morphology and velocity field of the Ly$\alpha$ nebula. We will then present appropriate models.
Here we describe an infall model that can explain the observations quite well. 
%The following coordinate system helps to visualize the geometry of the problem: the $z-$axis denote the L.O.S direction while the $x$ and $y$-axes denote the RA and dec-axes, respectively. 
An AGN sits in the center of a halo of mass $M_{\rm tot}=5 \times 10^{12}M_{\odot}$. A dusty torus surrounding the AGN causes it to photoionize a biconical region of the surrounding gas. The opening angle of each cone is assumed to be $\phi=90^{\circ}$. The cone axis is not perfectly aligned with the line-of-sight (L.O.S), but intersects it at an angle of $\sim 20^{\circ}$. Furthermore, the cones are rotated by $\sim 10^{\circ}$ counterclockwise on the sky. In this scenario, one cone is pointing almost directly at the observer, while the other points in the opposite direction. Photoionization and subsequent recombination in the cones converts ionizing radiation emitted by the AGN into Ly$\alpha$, and thus results in spatially extended fluorescent Ly$\alpha$ emission \citep{HR01}. In this picture, infalling gas in the nearest cone will be emitted with a redshift relative to the systemic redshift. 

We found that a model in which the gas density increases as $\rho\propto r^{-2}$, and the gas velocity increases $v(r) \propto r^{-1/2}$ can 
reproduce 
most observed properties of the nebula  (Fig.~13). In order to match the observations, it is crucial that the emission from the furthest cone is not observed (see below). For a more detailed description of the model, the reader is referred to Appendix~A.  According to Fig.~13, our infall model produces the following features:

\begin{itemize}

\item A rounded and 
centrally peaked morphology ({\it left panel}). This is a consequence of the geometry of our model
and the increase of the recombination rate (and thus of fluorescent Ly$\alpha$ emission) towards the AGN.
 The innermost contour level in Fig.~13 encloses a surface brightness that is $25$ times higher than that surrounded by the outermost contour. 
These properties are similar to those observed  (Fig.~ 2), within a similar nebular size. In order to obtain a more symmetric appearance, as shown
by the data, the axis cone should be closer to the line of sight. This would
correspond to a quasar.  The fact that  the AGN continuum and broad line
emission are detected at rest frame optical wavelengths, but are not obvious
in the UV rest frame could be explained by the existence of dust
obscuring the central regions. 

\item The largest redshift is observed at the Ly$\alpha$ centroid ({\it central panel}), as observed in the data
(Fig.~3, right). Because the infall velocity increases towards the AGN, the redshift of the Ly$\alpha$ line decreases outwards. Labels in the figure denote the mean velocity shift of the emission line in km s$^{-1}$ with respect to he centroid, i.e., the position of maximum Ly$\alpha$ flux. These values are consistent with the measured velocity offsets.

\item The nebular centroid has the largest redshift w.r.t the systemic redshift.
It is not possible to determine the systemic redshift in this object. 
However, it is interesting to note
that in MRC 1558-003 Ly$\alpha$ is redshifted  relative to
the main rest-frame UV   emission lines (CIV, HeII, CIII]) both in the spatially integrated
spectrum (R\"ottgering et al. \citeyear{rot97}) and at different  spatial
positions (Villar-Mart\'\i n et al. \citeyear{vm03}). In our scenario,
 the Ly$\alpha$ emission from
the far cone is negligible.
The other emission lines might also be fainter from the more distant
cone 
because
of extinction, but these effect would be less pronounced, since they are not affected by resonance scattering effects (see also Humphrey et al. \citeyear{hum07}).
As a result, 
the  relative contribution from the far cone to the flux of these lines  
is larger 
than for  Ly$\alpha$ both in the integrated spectrum and at different projected spatial
positions. Since the  emission from the more distant cone 
 must be blueshifted relative to the near cone because
of the infall pattern,  the 
lines will be blueshifted relative to Ly$\alpha$, which is consistent with the observations.

\end{itemize}

To reproduce these two last features, the Ly$\alpha$  emission detected from the furthest cone must
be negligible
as the expected blueshift of these  photons would eliminate the predicted increase of
 the Ly$\alpha$ redshift towards the nebular centroid and the blueshift relative
to other emission lines. Although radiation blueward of the rest-frame Ly$\alpha$ frequency is subject to absorption in the IGM, the IGM at $z=2.6$ is not opaque enough to completely eliminate the flux from the far cone (we used the model of Dijkstra et al. \citeyear{dijk07} to calculate the impact of the IGM on the Ly$\alpha$ line). Alternatively, the flux from the far side could be eliminated by a neutral and dusty spatially extended structure 
 that lies between the two cones. Several studies suggest the existence of
such structure (e.g. Humphrey et al. \citeyear{hum06}, van Ojik et al.
\citeyear{ojik97}).

Our model has greater difficulty reproducing the observed Ly$\alpha$ FWHM ({\it right panel}, Fig.~13). Especially if purely radial infall is considered, the model produces a maximum FWHM of $\sim$500 km s$^{-1}$, which is a factor of almost 3 short of what is observed for Ly$\alpha$. The model FWHM shown in Fig.~13 is boosted to $~750$ km s$^{-1}$ by convolving the spectrum at each pixel with a Gaussian with a standard deviation of $\sigma_{\rm 1D}=v_{\rm circ}/\sqrt{2}\sim 250$ km s$^{-1}$ ($\sigma_{\rm 1D}$ is the 1-D velocity dispersion of the halo). This reflects that the infall probably does not occur purely along radial paths. 
 Although inconsistent with the Ly$\alpha$ measurements,
this model reproduces successfully the spatial variation of the HeII FWHM  
observed by (Villar-Mart\'\i n et al. 2003) along the radio axis using long slit spectroscopic data. 
In this direction, the nebula shows broader FWHM$\sim$750 km s$^{-1}$
at the spatial centroid and $\sim$500 km s$^{-1}$ in the outer parts (Villar-Mart\'\i n et al. 2003),
consistent with the models.
 
 Resonant scattering of Ly$\alpha$ photons has been ignored in our model
because the gas  within the cones is highly ionized and  likely optically thin to Ly$\alpha$ (see Appendix~A). Although resonant scattering effects could explain the large Ly$\alpha$  FWHM values, and its broader profile compared with HeII (\S4.1), they would contradict other observational results.
 \citet{dijk06} have performed radiative transfer calculations through neutral collapsing gas clouds and found that resonant scattering does not only broaden the Ly$\alpha$ line, but that energy transfer from the infalling gas to the Ly$\alpha$ photons shifts the line to bluer wavelengths. On the contrary, the observations suggest that the Ly$\alpha$ line is redshifted with respect to other emission lines and probably relative to the system's redshift. Neither does resonant scattering through optically thick, collapsing, gas clouds reproduce the observation that the Ly$\alpha$ line is redshifted more towards the center. If resonant scattering does not occur, then it remains to be explained why Ly$\alpha$ shows such large FWHM values and it is 
 much broader than
 HeII. 
%Electron scattering of BLR Ly$\alpha$ photons might be an option. Given that the HeII line is 
%very weak in the broad line region of quasars (e.g. Heckman et al \citeyear{heck91}), this effect would not broa%den this is line.

An alternative  mechanism that can broaden Ly$\alpha$ relative to other emission lines
and could produce a similar velocity pattern  as that of infall
is outflows (Dijkstra, Haiman \& Spaans \citeyear{dijk06}).  In this scenario, a   shell propagates outwards
from the center, which is thick enough to be self-shielding (with column densities of hydrogen of 10$^{20}$ 
cm$^{-2}$). All  material is highly ionized, except inside the shell that is moving  outwards. In this case, Ly$\alpha$ photons basically bounce off the shell  as if it were a mirror, scattering back  and forth between the two expanding mirrors until they have been  Doppler boosted far enough for them to be able to simply propagate  through the shell (each time a Lya photon scatters off the shell, it  picks up a Doppler shift from the shell). This process will broaden Ly$\alpha$,
not affecting the other emission lines. Detailed modeling would be necessary to test whether outflows 
can reproduce all other observed properties of the nebula as infall does. However, since outflows cannot explain the radio/optical
asymmetries seen in MRC 1558-003 (Humphrey et al. \citeyear{hum07}), which on the other hand, are successfully
explained with infall, here we will not explore  the outflows scenario further.

In conclusion,  our infall  model  can
explain successfully the 
morphology, size, surface brightness distribution and the velocity field of the Ly$\alpha$ nebula associated with MRC 1558-003. It can also explain why Ly$\alpha$ is redshifted
relative to other emission lines and the FWHM values of the non resonant HeII line. The infall scenario is also consistent with our previous results
which imply that the {\it quiescent} nebulae in many HzRG (and MRC 1558-003 in particular)
are infalling  (Humphrey et al. \citeyear{hum07}).
On the other hand, this model  fails to explain the large Ly$\alpha$ FWHM
values. A mechanism which might not be connected with the infall process could be responsible
for the line broadening), while keeping the infall
pattern and the HeII FWHM intact. 

\vspace{0.2cm}

\centerline{\it MRC 2025-218}

\vspace{0.2cm}

It is not possible to investigate the kinematic pattern of the
quiescent gas, since the data do not allow to isolate its emission in the NE 
component.

\vspace{0.2cm}

\centerline{\it MRC 0140-257}

\vspace{0.2cm}

It is not possible to disentangle how the quiescent 
gas is moving in this object. The relaxed kinematics is not suggestive of outflows.

MRC 0140-257 
does not follow the trend found by Humphrey et al. (\citeyear{hum07}) in
powerful radio galaxies at   different  $z$, such that
    the quiescent gas has     the highest redshift
at the side of the brightest and more polarized radio hot spot.
The authors interpret these results as evidence for infall. On the contrary, 
the highest redshift in MRC 0140-257 is measured 
in $A1$, at the side of the fainter (northern) radio hotspot.

This discrepancy  could be due to three different
reasons: (i)  the orientation diagnostics based on the radio properties 
is not valid (ii) the gas is not in infall (iii)  our
 radio-optical registration (\S2) is incorrect.  If the radio core was located
on $A1$ or $A2$ (rather than between the two Ly$\alpha$ components),
we would have no information on the redshift on one side of the nucleus.

The interpretation of infall by Humphrey et al. (\citeyear{hum07}) 
 is  based on the fact  that the radio asymmetries in their sample  provide 
a powerful diagnostic tool to constrain the orientation of the radio source (and
the Ly$\alpha$ nebula as a consequence). One possible explanation 
for the MRC 0140-257 discrepancy is that  in this particular case
the radio 
properties are not reliable indicators of the object orientation.
This is actually  suggested by the fact that 
the brightest (southern) hot spot shows the lowest polarization (Carilli et al. \citeyear{car97}), contrary
to what we would expect if both flux and polarization asymmetries
 where mostly determined by orientation.

The orientation diagnostics based on the radio properties would 
fail, for instance, when the radio axis is very close to the plane of the sky.
For MRC 0140-257, on the contrary, the detection of a radio core  (Carilli et al. \citeyear{car97})
rather 
suggests that the angle between the radio axis and the plane of the
sky is not negligible.  

The diagnostics would also fail if  
the radio asymmetries are a consequence of environmental effects (e.g.
McCarthy, van Breugel \& Kapahi \citeyear{mac91b}), rather than orientation. 
This is possibly the case in MRC 0140-257.
On one hand, the brightest, less polarized hotspot is  closest to
the nucleus, and the Ly$ \alpha$ emission on that side of the nucleus (adopting
our present radio-optical registration) is also brightest.  This is what
we expect from McCarthy, van Breugel \& Kapahi (\citeyear{mac91b}) environmental scenario. On the other hand,
the southern (brighter) radio structure consists of two radio hot spots
 (Fig.8). If this is due to a  interactions with the gas 
(e.g. Carilli  et al. \citeyear{car97}), this process would enhance the
radio emission  and
 the higher brightness of the southern  radio structure
would therefore be a consequence of environmental effects.

Alternatively, it is possible  that the Ly$\alpha$ nebula is not infalling
towards the center.  $A1$ and $A2$ could be two objects rotating
around a common center of mass (e.g. De Breuck et al. \citeyear{breu05}).  If this is located between $A1$ and $A2$,
we estimate a
dynamical mass of $>$3$\times10^{9}$ M$_{\odot}$ for a radius $r$=16 kpc and a projected rotation velocity of 60 km s$^{-1}$ (i.e. half of the $V_{offset}$ value between $A1$ and $A2$). This mass value is at least a factor of 100
 smaller than  the stellar masses inferred for  other HzRG (Seymour et al.
 \citeyear{sey07}, Villar-Mart\'\i n et al. \citeyear{vm06}) or for massive, early-type
galaxies at $z\sim$2 in deep, IR surveys (e.g. Daddi et al. 2004).
 MRC 0140-257
could be an unusually small mass HzRG, a possibility that is also suggested
by the small FWHM values measured in $A1$ and $A2$. 
It would be essential to obtain  rest frame H and K magnitudes
(e.g. Seymour et al.
 \citeyear{sey07}, Villar-Mart\'\i n et al. \citeyear{vm06}) to make a proper estimation of the stellar mass.

If, on the contrary,
 this is progenitor of a massive elliptical
galaxy, as one expects for a powerful HzRG (e.g. McLure et al. \citeyear{lure99}), the rotation plane of $A1$ and $A2$ must almost
coincide with the plane of the sky.

\subsection{Ly$\alpha$ absorption}

Ly$\alpha$ absorption has definitively been detected in 
MRC 2025-218, spatially extended for $\ga$43$\times$32 kpc$^2$. The
absorber could  therefore be larger in size than the optical
size of the host (if already
formed) underlying galaxy.  CIV is also absorbed and, thus,
 the absorbing gas is highly ionized. As in this object, very little kinematic
structure along the line of sight was also
found by Wilman et al. (\citeyear{wil05}) across the giant
Ly$\alpha$ radio quiet nebula LAB-2  discovered by Steidel
et al. (\citeyear{ste00}). A more detailed analysis of the absorbers
associated with MRC 2025-218 will be presented in Humphrey et al. 
(2007, in prep.).

\section{Summary and conclusions}

The morphological and spectroscopic properties of the giant ($>$60 kpc) Ly$\alpha$ nebulae
 associated with three radio
galaxies (MRC 1558-003, MRC 2025-218 and MRC 0140-217) at $z\sim$2.5 have been investigated using VIMOS integral field spectroscopic data
on VLT.  The three objects  are associated with giant
($>$60 kpc) Ly$\alpha$ nebulae. This was already known for
MRC 1558-003 and MRC 2025-218, but not for MRC 0140-257.

The  morphologies are varied.
  In one source the nebula has a centrally
peaked, rounded appearance (MRC 1558-003), while in two objects
it consists of two spatial components (MRC 2025-218, MRC 0140-257). 
The total Ly$\alpha$ luminosities are in the range (0.3-3.4)$\times$10$^{44}$
erg s$^{-1}$.

The Ly$\alpha$ spectral profile shows strong variation through the nebulae, with FWHM
values in the range $\sim$400-1500 km s$^{-1}$ and velocity shifts $V_{offset}\sim$120-600 km s$^{-1}$.
 Kinematic disturbance
induced by the radio structures plays a clear important role in 
 MRC 2025-218.

Most spectroscopic
and morphological  properties  
of the giant nebula associated with MRC 1558-003 are successfully explained 
 by a scenario such that the giant nebula is collapsing
towards the center in the potential well of a 5$\times$10$^{12}$ M$_{\odot}$ halo. 
This is consistent with our previous conclusion that
the giant nebulae associated with this and  other HzRG are in infall
(Humphrey et al. \citeyear{hum07}). On the other hand, this model has problems to reproduce the large
Ly$\alpha$ FWHM  values ($>$1000 km s$^{-1}$), which might be a consequence of a mechanism unrelated
to the infall process.

We have discovered a   giant ($\sim$75 kpc) 
Ly$\alpha$ nebulae associated with a radio galaxy at $z=$2.64. It is
very closely aligned with the radio axis and it extends well beyond the radio
structures.
 It is
characterized by a strikingly relaxed kinematics (FWHM$<$300 km s$^{-1}$ and $V_{offset}\la$120  km s$^{-1}$), 
unique among HzRG. This object might have an unusually
small dynamical mass ($\ga$3$\times$10$^9$ M$_ {\odot}$) for a HzRG,
although this needs further investigation with other type of data. 

Ly$\alpha$ absorption is definitively detected in MRC 2025-218. The absorbing
screen has a spatial extension of 43$\times$32 kpc$^2$ at least.
It is highly ionized (CIV is also absorbed) and shows a remarkable velocity coherence
($<$100 km s$^{-1}$)
across its spatial extension.

\section*{Acknowledgments}
Thanks to an anonymous referee for useful comments on the paper. We  thank Chris Carilli and Laura Pentericci for
providing the radio maps presented in this paper.
The work of MV-M and RGD has been supported by the Spanish Ministerio de Educaci\'on y Ciencia and the Junta de Andaluc\'\i a through the grants AYA2004-02703 and TIC-114. SFS thanks the Spanish Plan Nacional de Astronom\'\i a, program
AYA2005-09413-C02-02 (MEC) and
the Plan Andaluz de Investigaci\'on of Junta de Andaluc\'\i a, research
group FQM322.
AH acknowledges support from a UNAM postdoctoral research fellowship.
Thanks to Joel Vernet for producing figures 3 and 8.

\appendix
\label{app}
\section{Infall Model of MRC 1558-003}

For each position $(\alpha,\delta)$ the spectrum was obtained by adding the emission from each cell $(\alpha,\delta,z)$ inside the cone along that line of sight (L.O.S). The emissivity for a cell was calculated by assuming photoionization equilibrium. This implies the volume emissivity at radius $r$ scales as $\epsilon(r)\propto n_H^2(r)$, where $n_H(r)$ is the number density of hydrogen nuclei at radius $r$, and we used that the gas is (almost) fully ionized.
 
In the frame of the cell (which itself is falling toward the center), the Ly$\alpha$ line is broadened by thermal motions of the atoms in the gas, which results in a Voigt profile. The bulk motions of the cell then cause a Doppler shift of the Ly$\alpha$ line. The total spectrum at $(\alpha,\delta)$ on the sky, denoted by $S(\alpha,\delta,x)$, is obtained by adding the contributions from all cells. To be more precise, the spectrum at $(\alpha,\delta)$ is given by
\begin{equation}
S(\alpha,\delta,x)\propto \int ds \hspace{1mm} \epsilon(\alpha,\delta,s) \hspace{1mm} \phi\Big{(}x-\frac{v(s)}{v_{\rm th}}\Big{)}
\label{eq:spectrum}, 
\end{equation} where $\epsilon(\alpha,\delta,s)= n^2_H(\alpha,\delta,s)$ if the cell is inside the cone, and otherwise $\epsilon(\alpha,\delta,s)=0$. Furthermore, $\phi(x)$ is the Voigt function\footnote{The Voigt function is given by $\phi(x)=\frac{a}{\pi}\int_{-\infty}^{\infty}\frac{e^{-y^2}dy}{(y-x)^2+a^2}$, where $a=A_{21}/4 \pi \Delta \nu_D$ $=4.7 \times 10^{-4}$ $(13 \hspace{1mm} {\rm km \hspace{1mm} s}^{-1}/v_{th})$. Here $A_{21}$ is the Einstein A-coefficient for the transition, $v_{th}$ is the thermal velocity of the hydrogen atoms in the gas, given by $v_{th}=\sqrt{2k_BT/m_p}$, where $k_B$ is the Boltzmann constant, $T$ the gas temperature, and $m_p$ the proton mass.}.

The gas density and velocity profiles were assumed to be of the forms $n_H(r) \propto r^{-2}$, $v(r)\propto r^{-1/2}$. The infall  velocity was assumed to be equal to the circular velocity of the halo at the  virial radius To prevent both fields from diverging, both $n_{H}$ and $v$ were kept constant at radii less than $0.15$ times the virial radius (which corresponds to $0.15\times 160=24$ kpc). These fields were chosen to give the best agreement with the observations. Note that these parameters were chosen to produce a good fit to the observations. To enhance the FWHM at each location $(\alpha,\delta)$ were convolved with a Gaussian of width $\sigma_{\rm 1D}=v_{\rm circ}/\sqrt{2}\sim 250$ km s$^{-1}$.

Equation~\ref{eq:spectrum} is valid in the limit where the cloud is transparent to all Ly$\alpha$ photons. In our model the Ly$\alpha$ from the furthest cone is not observed and we only need to consider Ly$\alpha$ emission from the nearest cone. For these photons we find that the optically-thin approximation is very reasonable: all Ly$\alpha$ photons are emitted in the cone where the gas is highly ionized. For the cone orientation of our model, each Ly$\alpha$ photon never leaves the cone on its way to the observer. The total optical depth for a given Ly$\alpha$ photon is reduced even further by velocity gradients along a L.O.S. Since all these photons are emitted predominantly redward of the Ly$\alpha$ resonance, scattering in the IGM is not important. 
%\lastpage

\end{document}